\documentclass[useAMS,usenatbib]{mn2e}
\usepackage{graphicx}
\usepackage{url}
\usepackage{xspace} %SdM - automatically fixes problems with spacing after \MSun symbol
\usepackage[dvipsnames]{xcolor} %SdM: colored comments

\def\aap{\ {A\&A}\ }
\def\aapr{\ {A\&A Review}\ }

\def\aaps{\ {A\&AS}\ }

\def\actaa{\ {Acta Astron.}\ }

\def\aj{\ {AJ}\ }

\def\apj{\ {ApJ}\ }
\def\apjl{\ {ApJL}\ }

\def\apss{\ {Ap\&SS}\ }
\def\araa{\ {ARA\&A}\ }

\def\memsai{\ {Mem. Soc. Astron. Ital.}\ }

\def\mnras{\ {MNRAS}\ }
\def\nar{\ {New Astron. Review}\ }
\def\nat{\ {Nat}\ }

\def\pasa{\ {Proc. Astron. Soc. Aust.}\ }
\def\pasp{\ {PASP}\ }

\def\physrep{\ {PhysRep}\ }

\def\pos{\ {Pub. of Science}\ }

\newcommand{\MSun}{\ensuremath{\,{\rm M}_\odot}\xspace}
\newcommand{\LSun}{\ensuremath{\,{\rm L}_\odot}\xspace}
\newcommand{\RSun}{\ensuremath{\,{\rm R}_\odot}\xspace}

\newcommand{\MSunyr}{\ensuremath{\,{\rm M}_\odot\,{\rm yr}^{-1}}\xspace}
\newcommand{\km}{\ensuremath{\,{\rm km}}\xspace}
\newcommand{\kms}{\ensuremath{\,{\rm km}\,{\rm s}^{-1}}\xspace}
\newcommand{\ergpers}{\ensuremath{\,{\rm erg}\,{\rm s}^{-1}}\xspace}

\def\apgt{\ {\raise-.5ex\hbox{$\buildrel>\over\sim$}}\ }
\def\aplt{\ {\raise-.5ex\hbox{$\buildrel<\over\sim$}}\ }
\def\lteq{\ {\raise-.5ex\hbox{$\buildrel<\over-$}}\ }

\title[Forming short-period Wolf-Rayet X-ray binaries and double black holes]{Forming short-period Wolf-Rayet X-ray binaries and double black holes through stable mass transfer}

\author[E.P.J. van den Heuvel, S.F. Portegies Zwart \& S.E. de Mink]{E.P.J. van den Heuvel$^{1,3}$ and S.F. Portegies Zwart$^{2}$ and S.E. de Mink$^{1,3}$\\
$^{1}$Astronomical Institute ‘Anton Pannekoek’, University of Amsterdam, 
P.O. Box 94249, 1090 GE, Amsterdam, The Netherlands \\
$^{2}$Leiden Observatory, Leiden University, PO Box 9513, 2300 RA, Leiden, 
The Netherlands\\
$^{3}$Kavli Institute for Theoretical Physics, UCSB Kohn Hall, Santa Barbara, CA 93106-4030, USA
}

\begin{document}
%\baselineskip=1.5\baselineskip

\date{}
%\pagerange{ -- } \pubyear{2011} 
\maketitle

\begin{abstract} %%<300 words

  We show that black-hole High-Mass X-ray Binaries (HMXBs) with O- or B-type donor stars and
  relatively short orbital periods, of order one week to several months may
  survive spiral in, to then form Wolf-Rayet (WR) X-ray binaries with orbital periods of
  order a day to a few days; while in systems where the compact star is a neutron star,
   HMXBs with these orbital periods never survive spiral-in. We therefore predict 
  that WR X-ray binaries can only harbor black holes.
  The reason why black-hole HMXBs with these orbital periods
  may survive spiral in is: the combination of a radiative envelope of
  the donor star, and a high mass of the compact star. In this case, when the donor
  begins to overflow its Roche lobe, the systems are able to 
  spiral in slowly with stable Roche-lobe overflow, as is shown by the system SS433. 
  In this case the transferred mass is ejected from the vicinity of the compact star
  (so-called ``isotropic re-emission'' mass loss mode, or ``SS433-like
  mass loss''), leading to gradual spiral-in. If the mass ratio of donor and black hole is $\apgt 3.5$, 
  these systems will go into CE evolution and are less likely to survive. 
  If they survive, they produce WR X-ray binaries with orbital periods of a few hours to one day.

  Several of the well-known WR+O binaries in our Galaxy and the Magellanic
Clouds, with orbital periods in the range between a week and several months, are
expected to evolve into close WR-Black-Hole binaries,which may later produce close
double black holes. The galactic formation rate of double black holes resulting from
such systems is still uncertain, as it depends on several poorly known factors
in this evolutionary picture. It might possibly be as high as $\sim 10^{-5}$ per year.

\end{abstract}

\begin{keywords}
stars: Wolf-Rayet stars, X-ray binaries, black holes, black-hole binaries
\end{keywords}

\section{Introduction}
Wolf-Rayet (WR) stars are hot and luminous evolved stars characterized by spectra with strong emission lines of He, C and/or N and O, produced by a dense high-velocity stellar wind. Their wind mass-loss rates are typically of order $10^{-5}$\MSunyr. Except for the most luminous WN stars, WR stars do not contain hydrogen; they are helium stars, as was first pointed out by \citet[][for recent reviews of WR star properties see \citealt{2007ARA&A..45..177C,2012A&A....540..144V,2016MNRAS.459..1505}]{1967AcA....17..355P}. WR X-ray binaries are composed of a helium star and a compact star, which can be a neutron star or black hole. Their existence was predicted by \citet[][]{1973NInfo..27...70T} and \citet[]{1973A&A....25..387V}, as the outcome of the later evolution of High-Mass X-ray Binaries (HMXBs), when the evolved massive O- or B-type donor stars in these systems started to overflow their Roche lobes. (The classical definition of a HMXB is: an X-ray binary in which the mass-donor star is an O- or early B-type star, while in a WR X-ray binary the donor is a massive helium star; we follow here this nomenclature). Van den Heuvel and DeLoore pointed out that the outcome of this later evolution of a HMXB is expected to be a very close binary, consisting of a helium star (the helium core of the original HMXB donor star) and a compact star. They suggested that the peculiar 4.8 hour orbital period X-ray binary Cygnus X-3 is such a WR X-ray binary. This was confirmed 19 years later \citep[]{1992Natur.355..703V}: the companion of Cyg X-3 is a WR star of type WN5. When the 1973 prediction of the existence of WR X-ray binaries was made, it was thought that all HMXBs would produce such systems. Since the WR phase lasts some 400 000 years \citep[][]{2015MNRAS..449..2322P}, which is longer than the duration of the HMXB-phase (e.g.\citealt{1994inbi.conf..263V}), one would expect that WR X-ray binaries would be more abundant in the Galaxy than the over 200 known HMXBs. However, apart from Cyg X-3, no other WR X-ray system has been found in the Galaxy, and the problem of the "missing WR X-ray binaries" \citep{1982A&A...115...69V,2005A&A...443..231L} has been with us for over 40 years. In recent years six more WR X-ray binaries have been discovered in other galaxies \citep[]{2015MNRAS..482..1112D}; with the exception of one, all of these have short orbital periods, between 0.2 and 1.5 days; one system has a period of about 8 days. Two possible factors that may lead to the reduction of the predicted numbers of WR X-ray binaries in the Galaxy are:
\begin{enumerate}
\item[(1)] when the existence of WR X-ray binaries was predicted it was thought that all helium  
stars with masses $\apgt 3 - 4$\MSun would be observable as WR stars. However, more recent estimates of
masses of WR stars, either in binary systems or from their absolute luminosities have shown that, in order to show the typical WR spectral characteristics produced by a dense high-velocity stellar wind outflow, the helium stars must have a mass $\apgt 10 (\pm2)$\MSun
\citep{2007ARA&A..45..177C,2012A&A....540..144V}. This implies that the hydrogen-rich
progenitor of the WR star most probably must have had a mass $\apgt 30
(\pm5)\,\MSun$, and in any case $\apgt 20\MSun$. 

Or:
\item[(2)] Only very few HMXBs survive the spiral-in during the Roche-lobe overflow phase that follows the HMXB phase (we use here the expression "spiral-in" for a drastic decrease of the orbital period by any kind of mechanism, not only by Common-Envelope (CE) evolution).
\end{enumerate}

Here we will argue that the latter factor is the key reason for the scarcity of WR X-ray binaries, and that HMXBs in which the compact star is a neutron star hardly ever survive spiral-in. (Only very wide neutron-star systems with donor masses below about $\ 20\MSun$ may survive, see section 3).  
  
We show that, on the other hand, if the compact star is a black
hole, with a mass $\apgt 5-10\MSun$, the systems may survive
spiral-in, and become close binaries consisting of a WR star and a
black hole. For this reason, one expects in general that in WR X-ray
binaries the compact star is a black hole.

Our ideas about which systems may survive spiral-in and produce WR X-ray binaries were triggered by the realization that the peculiar X-ray binary SS433 has avoided going into Common-Envelope (CE) evolution, and that the donor star in this system is transferring mass to the compact star by stable Roche-lobe overflow \citep{1999ApJ...519L.169K,2000ApJ...530L..25K}. By analysing the properties of this system we realized that it is the high mass of the compact star in this system ($4.3(\pm 0.8)\MSun$) in combination with the relatively low mass of its donor star ($12.3(\pm3.3)\MSun$) \citep[]{2008ApJ...676..L37H}, which allowed it to avoid Common-Envelope (CE) evolution and enables it to gently spiral-in without ever coalescing with its donor. As we consider SS433 to be a "keystone" for understanding the formation of the WR X-ray binaries, we give in section 2 a brief overview of its properties and evolutionary state. The avoidance of its going into CE-evolution is - as we will argue - a consequence of the donor star having a radiative envelope \citep{2000ApJ...530L..25K}, in combination with the donor star and accretor having a mass ratio less than 3.5. In section 3 we then examine for which donor masses, mass ratios and orbital periods HMXBs will, when they start Roche-lobe overflow, avoid going into CE-evolution and may survive as WR X-ray binaries with short orbital periods. We also examine under which conditions they may still survive after having gone into CE-evolution.   
In this section some examples are given of
how a number of well-known observed WR+O binaries with relatively
short orbital periods are expected to evolve in the future, and are
expected to produce WR X-ray Binaries and, as a final evolutionary state, close double black holes. In section 4 we attempt to estimate the birthrate of WR X-ray binaries in the Galaxy on the basis of our model, and find it to be still higher than observed and discuss possible ways to relieve this discrepancy. In section 5 we discuss the results and estimate the possible birthrate of double black holes based on our model.

\section{SS433 - a keystone for understanding which HMXBs go into stable Roche-lobe overflow}

\subsection{The evolutionary state and future evolution of SS433}

The SS433 system consists of a Roche-lobe filling A4-7I supergiant
donor star with an estimated mass of $12.3 (\pm 3.3)\MSun$ and a
luminosity of about 3800\LSun, plus a compact star with a mass of $4.3
(\pm 0.8)\MSun$, in a 13.1-day period binary
\citep{2008ApJ...676..L37H}. The compact star is surrounded by an
extended and luminous accretion disk, about an order of magnitude
brighter than its A-supergiant companion. This disk ejects the famous
precessing relativistic jets with a velocity of 0.265c, in which
neutral hydrogen is ejected at a rate of some $10^{-6}$\MSunyr,
while in a strong disk wind with a velocity of about 1500\kms of
order some $10^{-4}$\MSunyr is ejected, as is seen in the form of the
‘stationary’ H$\alpha$ line and broad absorption lines
\citep{2004ASPRv..12....1F}. The total mass loss from the disk is
basically all the matter that the A-supergiant donor is transferring
to the compact object by Roche-lobe overflow on its thermal timescale
of $\sim 10^5$ years \citep[see also][]{2006MNRAS.370..399B}.  

The observed radiative accretion luminosity of the compact star with its
disk does not exceed the Eddington luminosity $L_{\rm Edd} \simeq 6
\cdot 10^{38}$\ergpers of the compact star (which corresponds to a real
accretion rate onto the compact star of order only a few times
$10^{-8}$\MSunyr), although when seen along the jets the UV
luminosity might be as large as perhaps $10^{40}$\ergpers
\citep{2004ASPRv..12....1F}, which would correspond to an accretion
rate of order $10^{-7}$\MSunyr . This mass loss has been going on for
thousands of years, as can be seen from the large radio nebula W50
that surrounds the system and has been produced by the precessing jets
and the strong disk wind. Even though this mass transfer has been
going on for thousands of years, the system has not entered in a
Common-Envelope (CE) state.

The reason why SS433 has not gone into a CE phase is, as argued by
\citet{1999ApJ...519L.169K} and \citet{2000ApJ...530L..25K}, the fact that the
A-supergiant star has a radiative envelope. If one takes away mass
from a star with radiative envelope, this envelope responds by shrinking on a dynamical timescale, followed by a re-expansion on the thermal timescale of the envelope. As a result, this star can keep its
radius close to that of its Roche lobe and will transfer matter to its
companion on the thermal timescale of its envelope, without going into
a CE phase. There is, however, an extra condition for keeping the Roche-lobe overflow stable, which was not mentioned in the above references, namely: the
thermal timescale mass-transfer from a radiative donor envelope may
itself become unstable, if the mass ratio of donor and companion star
is larger than a value in the range 3 to 4 \citep[e.g.][]{1997MNRAS.291..732T,2000ApJ...530L..93K,2000MNRAS.315..543H}. 

For the sake of argument we will assume here this limiting mass ratio to be 3.5. For mass ratios larger than about 3.5 the 
shrinking of the system due to the mass transfer goes so fast that the shrinking of the donor star cannot keep in pace with it, and the system will enter a CE phase.  
SS433 has indeed a mass ratio below 3.5 and therefore avoided going into a CE phase.
Apart from systems with a donor with a radiative envelope 
and mass ratio larger than 3.5, also for systems in which the donor has a convective envelope,
the formation of a Common Envelope is unavoidable. The reaction of a convective 
envelope to mass loss is expansion on a dynamical (pulsational) timescale, 
and thus the envelope becomes violently unstable, which leads to runaway mass
transfer and the formation of a Common Envelope.

\subsection{Why systems like SS433 are so rare: the fate of HMXBs 
  with a ``standard'' neutron star companion}
  
  One may wonder what is so special about SS433 and why we do not see
  more SS433-like systems. We propose that the answer is: the very
  unusual combination for a HMXB of a rather low donor mass (presently
  $\sim 12.3$\MSun and initially $\sim 14$ to $\sim15$\MSun) plus a quite
  massive compact star ($\sim 4.3$\MSun). An 14 to 15\MSun initial
  donor mass and an orbital period of $\sim 13.1$\,days are typical for
  a Be/X-ray binary, the most common type of HMXB, containing a B-emission (Be) line star. There are some 200 Be/X-ray binaries known in our Galaxy and the Magellanic Clouds. 
  \citet{2005A&AT...24..151R} list 160 in our Galaxy and the two
  Magellanic Clouds and \cite{2011Ap&SS.332....1R} lists 141 in our
  Galaxy plus the SMC alone. Since these papers appeared, Swift,
  INTEGRAL and other satellites have discovered several tens more,
  bringing the total presently known number to about 200. 
  In all but one of the known Be/X-ray binaries, the compact stars are
  neutron stars which have a typical mass of about 1.4\MSun. Only
  one Be/X-ray binary is known to harbor probably a black hole
  companion, with a mass of 3.6 to 6.9\MSun
  \citep{2014Natur.505..378C}. If the companion of a 14\MSun Be star
  is a 1.4\MSun neutron star, the mass ratio is 10, and the formation 
  of a Common Envelope is unavoidable, and the two stars will merge (unless the orbital period is longer than 1 to2 years, which is the case for only a small fraction of the Be/X-ray binaries,  see section 3). 
  
  Only if the compact star has a mass $\sim 4$\MSun or larger, and the 
  donor has a radiative envelope, the system will spiral in
  slowly and survive the SS433-like mass-transfer process. Therefore,
  out of the $\sim 200$ Be/X-ray binaries, only the one system with an
  (alleged) black hole companion will in the future evolve like SS433
  and survive, all the others will have merged after transferring only
  a small amount of mass (or, the small fraction of systems with orbital periods longer than about 1 to 2 years, will have produced a very close Helium star plus neutron star system, after a short-lasting Common-Envelope phase). So, the birthrate of SS433-like systems is at most about 0.5 per cent of the birthrate of Be/X-ray binaries.  \citep[For a
    model for the formation and future evolution of the possible Be/black-hole binary,
    see][]{2015MNRAS.452.2773G}.

  The simple reason why SS433 can stably survive this type of
  spiral-in process for $\sim10^4$\ to perhaps $\sim10^5$\ years is because of its unique
  combination -- for a Be/X-ray binary progenitor -- of a quite massive
  compact star and a relatively moderate-mass donor star.
  
  The $\sim 4.3 $\MSun compact object in SS433 must be a low-mass
  black hole, because causality allows neutron stars to have masses
  not larger than about 3\MSun
  \citep{1973ApJ...179..277N,1996ApJ...470L..61K}.

\section{Conditions for survival of Roche-lobe overflow in HMXBs as a function of donor mass, compact star mass and orbital period}

\subsection{Evolution of the orbit during SS433-like mass transfer}

As shown by \cite{2000ApJ...530L..25K} and \cite{2006MNRAS.370..399B}, in the case of Roche-lobe
overflow from donor stars with a radiative envelope, a further
condition for avoiding the formation of a Common Envelope is that the
spherization radius $R_{\rm sp}$ of the accreting compact object
remains smaller than its Roche lobe, where $R_{\rm sp}$ is given by
\citep{1973A&A....24..337S}:
\begin{equation}
  R_{\rm sp} = - {27\over 4} {{\dot M}_{\rm donor} \over {\dot M}_{\rm Edd}} R_{\rm s},
\end{equation}
where $R_s$ is the Schwarzschild radius of the compact object. In the
case of SS433, $-{\dot M}_{\rm donor}/{\dot M}_{\rm Edd}$ is of order
$10^4$ and $R_{\rm s} \simeq 9$\km, so $R_{\rm sp} \sim 6 \cdot 10^5$\km
$\simeq 0.9$\RSun, which is deep inside the Roche lobe of the
compact star, and a CE will be avoided. In all HMXBs with
orbital periods upward on one day the same will hold. So in all cases
of HMXB systems with a donor with a radiative envelope and mass ratio less 
than about 3.5, one expects the system to go into normal Roche-lobe overflow 
evolution similar to that of SS433. The ``SS433-mode'' of mass transfer is 
what we have in the past called ``isotropic 
re-emission''\citep[e.g.][]{1991PhR...203....1B,1997A&A...327..620S,1994inbi.conf..263V,2006csxs.book..623T, 1975MmSAI..46..217M}.

With the SS433-mode of mass transfer, followed by mass loss from the
disk, which has the specific orbital angular momentum of the compact
object, it is simple to calculate how the orbit of the system will
change. In case that a fraction β of the transferred matter is ejected
from the compact star and its disk with the specific orbital angular
momentum of this star, and a fraction $(1-\beta)$ is accreted by this
star, the orbital angular momentum loss leads to a change of the
orbital radius a given by
\citep[e.g. see][]{1996A&A...315..453T,1997A&A...327..620S,2006csxs.book..623T}:
\begin{equation}
  a/a_o = {q_0 +1 \over q+1} \left( {q_0 \over q} \right)^2
  \left[ {(1-\beta)q_0+1 \over (1-\beta)q+1} \right]^{-3-2/(1-\beta)},
\end{equation}
where q is the mass ratio of donor and compact star, and subscript
zero indicates the initial situation at the onset of Roche-lobe overflow.

For the case in which in which $\beta = 1$, as is in fact the case in
SS433, as the accreted amount is $10^{-4}$ to $10^{-3}$ times the
transferred amount, this equation in the limit of $\beta$\,
approaching unity simplifies to:
\begin{equation}
  a/a_0= \left( {q_0+1 \over q+1} \right) \left( {q_0 \over q} \right)^2 e^{-2(q_0-q)}
\end{equation}
Using Kepler’s third law the corresponding equation for the change of the orbital period is:
\begin{equation}
  P/P_0= \left( {q_0+1 \over q+1} \right)^2 \left({q_0 \over q} \right)^3 e^{-3(q_0 - q)}.
\end{equation}

  In the case of SS433, assuming the initial mass of the A-supergiant
  donor to have been $\sim 14$ to $15\MSun$, the mass of its helium
  core is about 3.5\MSun. This means that at the end of the Roche-lobe
  overflow phase $q = 0.81$, while at present $q_0= 2.86$. Inserting
  these values into equation (4), one finds that at the end of the
  Roche-lobe overflow the orbital period of the system will be $P
  \simeq 5.60$ days. So, SS433 will with these assumed component
  masses finish as a detached binary consisting of a 3.5\MSun
  helium star and a 4.3\MSun compact star.
  The entire process will take place on the
  thermal timescale of the envelope of the 10\MSun A-supergiant
  which is between $\sim 10^4$ and $\sim 10^5$ years.

  The helium star in the resulting system may during helium shell
  burning go through a second mass-transfer phase and finally explode
  as a supernova, likely leaving a neutron star. If the system remains bound in response to the natal kick of the neutron star, a close eccentric  binary will result, consisting of the present $\sim 4.3$\MSun
  compact star plus a neutron star.

  \subsection{Upper limiting orbital period for having a radiative envelope \label{uplim_rad}}
In order to determine the limiting orbital period for having a
radiative envelope, we notice that for an effective temperature
$T_{\rm eff}>8100$\,K (spectral type earlier than 
$\sim$A7), stars have a deep radiative envelope
\cite[e.g. see][]{1968psen.book.....C}. If the luminosity of a donor
star of a given mass M is known, its radius $R$ can be found from the
relation $L= 4\pi \sigma R^2 T_{\rm eff}^4$, where $\sigma$ is the 
Stefan-Boltzmann constant.  If the mass $M_{\rm c}$ of the
compact companion of the star is known, and we set the stellar radius
$R$ equal to the radius $R_{\rm L1}$ of the Roche lobe of the donor, then
the equation for the Roche lobe given by \cite{1983ApJ...268..368E}:
\begin{equation}
  R_{\rm L1} = {0.49 \, a \over 0.6 + q^{-2/3} \ln{ (1+q^{1/3})}},
\end{equation}
allows one to calculate the orbital radius $a$ of the binary in which
the donor fills its Roche lobe, as the mass ratio $q=M_{\rm donor}/M_{\rm c}$ is
known. Using Kepler's third law, one also calculate the corresponding
orbital period $P$. 

\begin{figure}
\begin{center}
\includegraphics[width=0.5\textwidth]{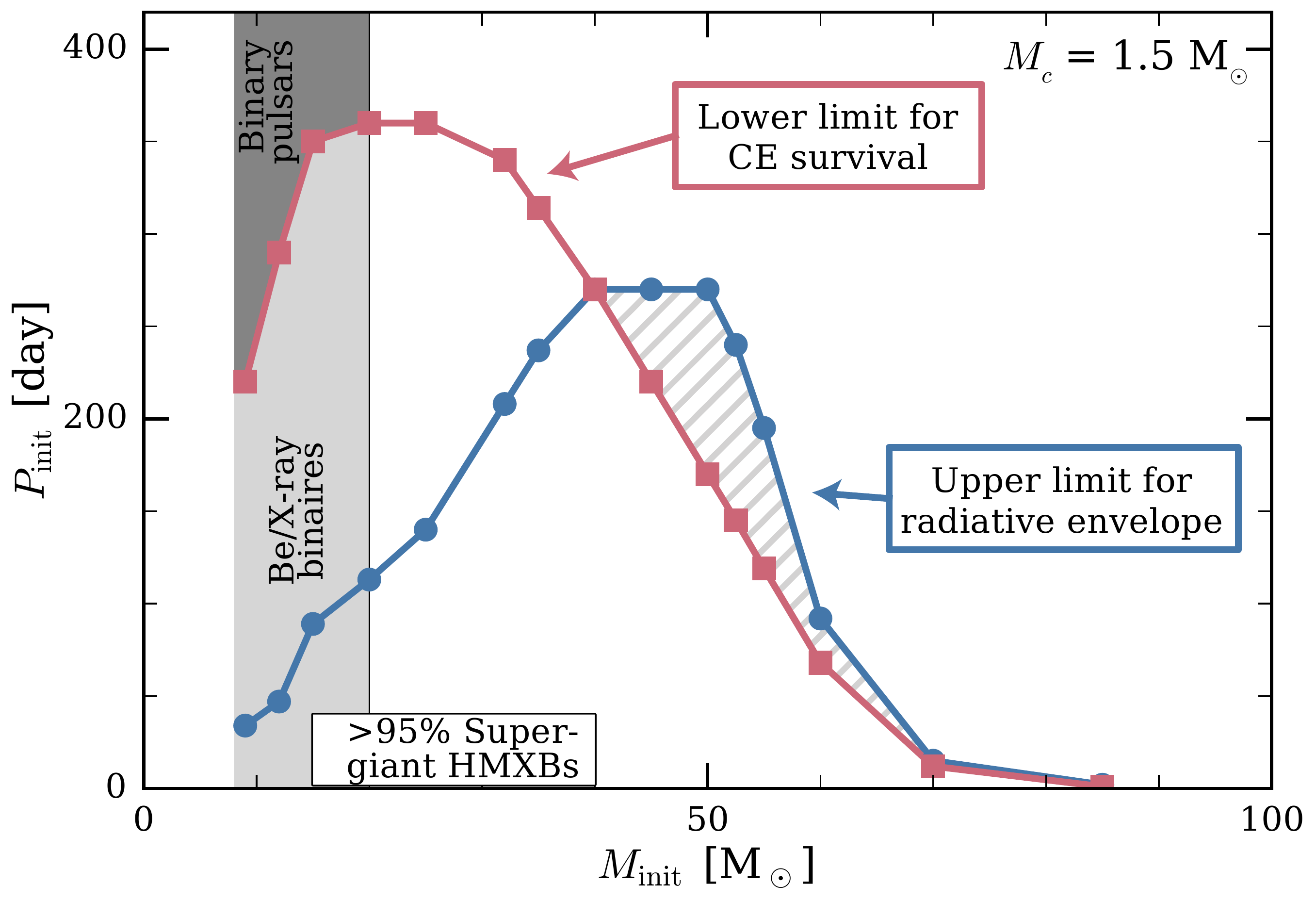}%
\end{center}
\caption{Upper limiting orbital periods for having a donor star with a radiative envelope  (blue line, circles, as explained in section~\ref{uplim_rad}), together with the formal lower limiting orbital periods for surviving CE evolution (red line, squares, as explained in section~\ref{uplim_CE}).  This diagram shows the case of a 1.5\MSun neutron star companion (see figure \ref{Fig:2} for more massive companions). In all cases considered here the mass ratio between donor and accretor is so large ( $q > q_{\rm limit} = 3.5$) that Roche-lobe overflow is expected to be unstable and lead to CE evolution, irrespective of whether the donor has a radiative envelope. Only systems with periods larger than the limiting period for CE-survival (above the red curve) are expected to survive spiral in and avoid coalescence. A small region of the parameter space allows for radiative donors that may survive CE inspiral (gray hashed region), but this is limited to donors with masses above 40\MSun. The typical orbital periods and donor masses of the observed supergiant HMXBs and the Be-HMXBs are indicated. Over 95\% of the NS-supergiant HMXBs do not survive
  SS433-like spiral in, and only the Be-HMXBs with very long orbital periods survive CE evolution and can be progenitors of binary pulsars.
\label{Fig:1}}
\end{figure}

\begin{figure}
\begin{center}
\includegraphics[width=0.5\textwidth]{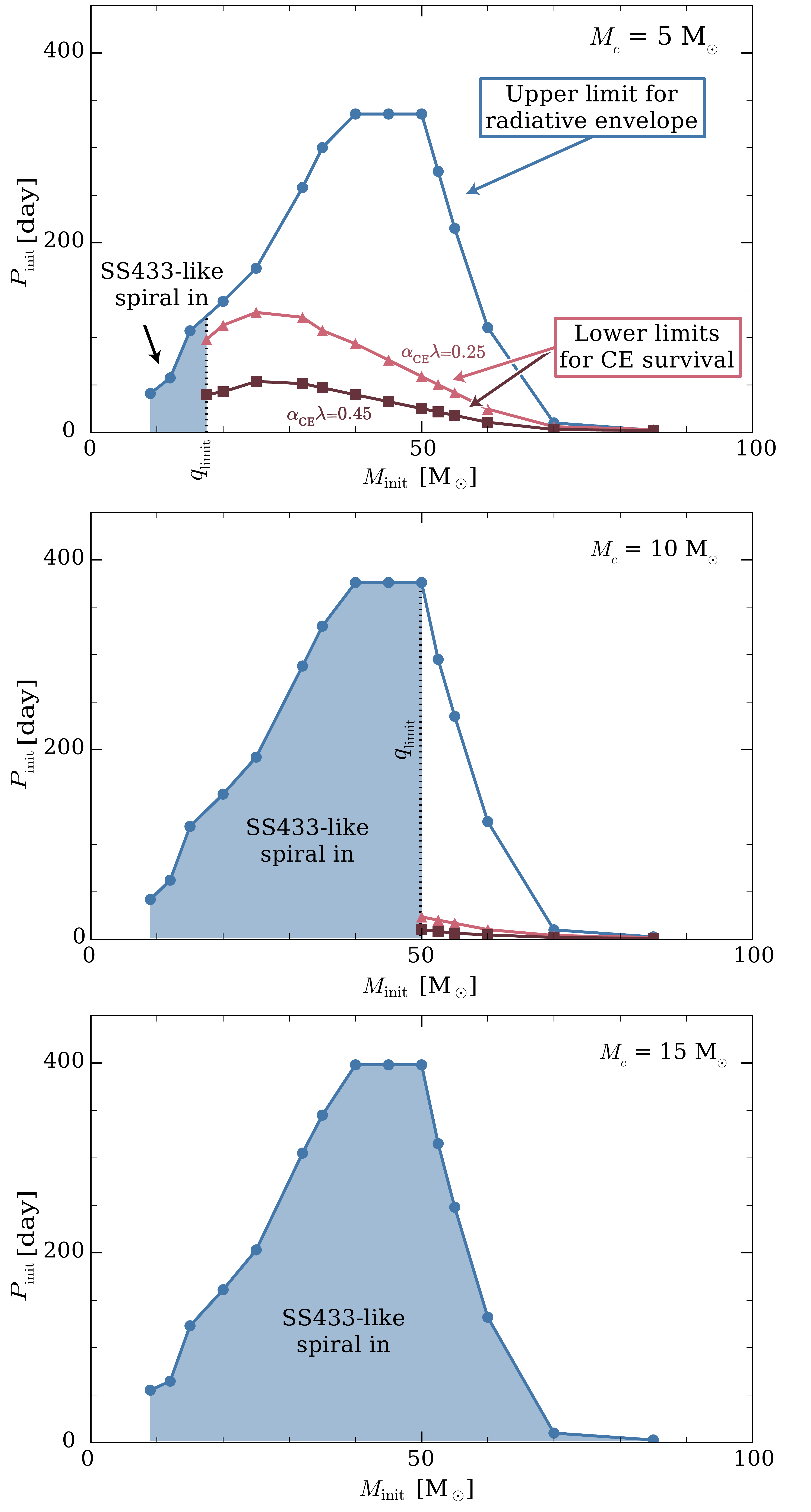}
\end{center}
\caption{Blue shaded region indicates systems that are expected to undergo stable mass transfer from a radiative donor and survive a ``SS433-like spiral in''. Top, middle and bottom panel show the range for a black-hole companion with a mass $M_c = 5$, $10$ and $15\MSun$, respectively. (Compare with figure~\ref{Fig:1}, where we showed the case of a $M_c = 1.5$\MSun neutron star companion). The region is bounded by the condition that the mass ratio between the donor and the compact object is not too extreme, i.e.\ smaller than $q_{\rm limit} =3.5$ (vertical dashed gray line). Systems in the part to the right of the blue region will go into CE evolution.
The upper limiting period for mass transfer from a donor with a radiative envelope (blue line with circles, see section~\ref{uplim_rad}) and the the lower-limiting orbital-period curves for survival of CE evolution as WR X-ray binaries are shown for two values of the CE parameters (light and dark red lines, see section~\ref{uplim_CE}). 
\label{Fig:2}}

\end{figure}

To calculate the maximum orbital periods up to which
donor stars still have a radiative envelope, we used the luminosities
of post-main-sequence evolutionary tracks, for solar metallicity, of
rotating stars with masses up to 50\MSun given by
\cite{2012A&A...537A.146E}, and the limiting effective temperature of
$T_{\rm eff} \sim 8100$\,K. Post-main-sequence
stars originating from stars more massive than about 50\MSun have very strong stellar wind mass loss, or become Luminous Blue Variables, stars that experience strong eruptive mass loss episodes. Because of their strong mass loss they lose most of their hydrogen-rich envelope and always stay at effective temperatures above 8100K. Therefore these post-main sequence stars are expected to have
radiative envelopes, so for them we used as maximum radius just their
maximum post-main-sequence stellar radius. We made these calculations
for compact companions with masses of 1.5\MSun, 5\MSun, 10\MSun
and 15\MSun. Figures~\ref{Fig:1} and \ref{Fig:2} %, \ref{Fig:4} and \ref{Fig:5}
give these upper limiting orbital periods as a function of donor mass
for initial donors with masses between 9\MSun and 85\MSun. (In
the mass range between 40\MSun and 60\MSun there are no tracks
by \cite{2012A&A...537A.146E} available. It is known from evolutionary
calculations with similar assumed wind mass loss rates that for masses
above 50\MSun the stars at the end of hydrogen burning have lost
most of their H-rich envelopes and their radii drop rapidly as a
function of mass. We have, for the sake of argument, assumed that in
the mass range between 40 and 50\MSun the orbital periods at $T_{\rm eff} = 8100$\,K are
constant and after that they linearly decrease towards the orbital period of
the 60\MSun star). 

The figures show that for donor masses up to
about 50\MSun these upper limiting orbital periods range from
about 50 to 400\,days, and beyond 50\MSun they go down
rapidly. Below these limiting curves, the donor stars in HMXBs will
transfer mass to their compact companions according to the SS433-type
of mass transfer, and the systems will not go into Common Envelope
(CE) evolution, provided the mass ratio of donor and compact star is
less than about 3.5. The regions where this SS433-like 
evolution will occur are indicated in the figure 2 by the blue-colored 
parts of the diagrams.(Notice that for calculating the curves of the limiting orbital periods, as well as the limiting donor masses for mass ratio 3.5, we used the real post-main-sequence masses of the stars, which are considerably reduced with respect to the intitial masses, due to stallar wind mass loss on the main sequence). %EdvdH: I  added this sentence between parentheses for clarity   
To the right of these blue regions and below the 
radiative boundary periods, systems will go into Common Envelope evolution 
with a donor with radiative envelope. Above the radiative boundary periods 
they will go into Common Envelope evolution with a convective envelope (or if the period is too large they will not experience mass transfer at all).

The next question is: which systems with donor stars in this
radiative-envelope regime, will survive as binaries after the onset 
of mass transfer by Roche-lobe overflow? It turns out
that none of the systems with a 1.5\MSun compact star (neutron
star) will, as was already mentioned above. Because of their
mass ratios of far above the mass ratio upper limit of 3.5, 
they go into CE evolution and for initial donor masses below 40 \MSun 
they all merge. This can be seen from the lower-limit curve for survival 
of CE evolution (the red curve) in figure~\ref{Fig:1}. (The way this curve was 
calculated is explained in the next section). Although for donor masses 
larger than 40\MSun according to the figure systems might survive CE-evolution (grey hashed region), such massive donors with NS companions may not be very likely,
for stellar evolution reasons. % EdvdH: I tuned this statement to be less strong. 
In figure~\ref{Fig:1} we have
indicated also the ranges of orbital periods and masses of the bulk of
the known supergiant HMXBs with neutron star companions, and of the
B-emission X-ray binaries with neutron star companions. One observes
that the vast majority of the known supergiant HMXBs with neutron
stars does not survive spiral-in.  

It can be seen from this figure that
among the known types of HMXBs only the B-emission X-ray binaries with
a neutron star companion and orbital periods ranging from larger than 220\,d with a
9\MSun donor to larger than 370\,d with a 20\MSun donor (the red region of the diagram), 
will survive spiral in and can later form double neutron stars, after explosion of the
helium star. This is a well-known result \citep[e.g. see][]{1996IAUS..165....3T,1996A&A...312..670P,2005MNRAS.363L..71D}.

While none of the systems in the radiative-donor regime with a 1.5\MSun neutron star 
survives the mass transfer, the systems with a 5\MSun, 10\MSun and 15\MSun compact
companion star (black hole), in the blue-colored parts of figure\,\ref{Fig:2}, do survive 
SS433-like mass transfer and can produce helium-star binaries with relatively short 
orbital periods: WR/X-ray binaries. 

It should be kept in mind that for the case with a 5\MSun, 10\MSun and
15\MSun compact star, the donors which we consider, overflow their
Roche lobes after leaving the main sequence, which means that they
evolve according to case B of close binary evolution \citep{1990sse..book.....K}. The lower limit
for the orbital period for this case is about one week.

An example of a system that will survive SS433-like evolution is Cygnus X-1,
which has 5.6 day orbital period, a $14.8\pm 1.0\MSun$ black hole and a
$19.2\pm1.9\MSun$ donor, which according to its luminosity must have started
out as a 30\MSun star \citep{2011ApJ...742...840}. After SS433-like mass
transfer the donor in Cygnus X-1 will leave a helium star of about
10\MSun, which might either leave a neutron star or a low-mass black
hole, with an orbital period of about 9 days, depending on the direction and magnitude of the birth kick of the compact object. So Cyg X-1 will not terminate as a close system.

\subsection{Lower limiting orbital periods for surviving CE evolution\label{uplim_CE}}

Systems with orbital periods above the upper-limit line for radiative envelope in figures~\ref{Fig:1}
and ~\ref{Fig:2} will have convective envelopes and will, when their donor stars
overflow their Roche lobes, go into CE evolution and spiral in, 
%in the usual way 
as described e.g. by \citet{1993MNRAS.260..675T}, \citet{1996A&A...309..179P}, \citet{1997AstL...23..492L} and
later papers and recently by \citet{2016Natur.534..512B} and \citet{2016MNRAS..462..3302E}. 

The same holds for systems 
with donors in the radiative region and mass ratios larger than about 3.5.
We use here the
formalism for the orbital change in the case of CE evolution as given
 by \cite{1984ApJ...277..355W} and \cite{1990ApJ...358..189D}, which
yields a ratio of the final and initial orbital radii $a_{\rm f}$ and $a_{\rm i}$,
respectively, given by:
\begin{equation}
  a_{\rm f}/a_{\rm i} = { M_{\rm core}\, M_{\rm c}/M_{\rm donor} \over M_{\rm
      c} + 2M_{\rm env}/(\alpha_{\rm CE} \, \lambda\, r_{\rm L})}
\label{Eq:af_ai}\end{equation}
where $M_{\rm core}$ is the mass of the helium core of the donor,
$M_{\rm env} \equiv M_{\rm donor} - M_{\rm core}$ is the mass of 
the hydrogen-rich envelope of the donor
at the moment when Roche-lobe overflow begins, 
$\alpha_{\rm CE}$ is
the so-called ``efficiency factor'' of CE evolution which indicates
the efficiency with which the release of orbital gravitational binding
energy that occurs during spiral in of the compact star towards the
core of the donor is converted into kinetic energy required to eject
the common envelope, $\lambda$ is a parameter that depends on the
stellar mass density distribution and $r_{\rm L}$ is the ratio of the
Roche-lobe radius $R_{\rm L}$ and the orbital radius $a_i$ at the onset
of CE evolution. 

The value of $r_{\rm L}$ is typically of order
0.5. There are many factors that influence the precise value of the product
$\alpha_{\rm CE} \lambda$, such as energy sources like recombination energy 
and accretion  energy release (e.g. see the discussion in
\citealt{2006csxs.book..623T,2013MNRAS.429L..45P}, and in \citealt{2013A&ARv..21...59I,2016A&A..596...A58K}). We assume 
here two values of the common envelope efficiency: $\alpha_{\rm CE} = 0.5$ and $\alpha_{\rm CE} =0.9$
(e.g. \citealt{1996IAUS..165....3T}), $r_{\rm L} =0.5$, and - while we know this 
an oversimplification - for our calculations we assume  $\lambda = 0.5$. 

To calculate the outcome of CE evolution for stars in the mass range
10 to 85\MSun we used the rotating evolutionary models for solar
metallicity of \cite{2012A&A...537A.146E}. These models were
calculated with stellar wind mass loss, such that at the end of
core-hydrogen burning, when the stars leave the main sequence, their
masses are lower than their initial values, and the mass of the helium
core is known. Using for $M_{\rm donor}$ then these reduced
post-main-sequence masses, and the corresponding $M_{\rm core}$
values, one can calculate, for given values of $M_{\rm c}$ and of the
combination $\alpha_{\rm CE} \lambda r_{\rm L}$, what the values of the
ratio $a_{\rm f}/a_{\rm i}$ will be. 

Given the mass $M_{\rm core}$ of the helium
core one knows the radius of this helium star (for these we used the interpolation 
formula from Onno Pols given in \citealt{2006csxs.book..623T}). Together with the mass
$M_{\rm c}$ of the compact star this radius gives the minimum
final orbital radius $a_{\rm f}$ for systems that survive CE-evolution, as the
helium star is not allowed to be larger than its Roche-lobe radius in
the final system. If $a_{\rm f}$ would be smaller than this minimum value, the
system does not survive and merges.

  Using the above-given values of $\alpha_{\rm CE} \lambda$ and $r_{\rm L}$,
and the values of $a_{\rm f}/a_{\rm i}$ calculated according to the above given recipe
for each initial donor mass and $M_{\rm c}$, one can then calculate
the lower-limits to the initial orbital radii $a_{\rm i}$ required for the
systems to survive CE evolution. These $a_{\rm i}$ values give one also the minimum initial
orbital periods for surviving CE evolution, as a function of initial
donor mass and $M_{\rm c}$.

In figure~\ref{Fig:1} the lower limiting period curve for survival of CE
evolution for $\alpha_{\rm CE}= 0.5$ is indicated for systems with a neutron star companion with
$M_{\rm c} = 1.5$\MSun. 
%As mentioned above, the part of the curve for donor masses larger than 40\MSun is not expected to be relevant, as in such massive systems the first-born compact object will not be a neutron star.
%
For the case of $M_{\rm c} = 5$\MSun, 10\MSun and 15\MSun, the
calculated lower limiting period curves for CE evolution with the
above-given values of $\alpha_{\rm CE} \lambda = 0.25$ and 0.45 are 
given. (The latter value was derived from orbits of post-CE binaries
by \citealt{1996A&A...309..179P}). Systems above these curves in the parts of the panels in  figure~\ref{Fig:2}, to the right of the blue-colored regions are expected to survive CE-evolution  
as WR/X-ray binaries with short orbital periods. 

\subsection{Examples of the future evolution of some well-known WR+O spectroscopic binaries:
  formation of close WR X-ray binaries and of double black holes}

Table~\ref{Table1} shows as an example how we expect seven well-known
observed massive WR+O spectroscopic binaries with well-determined
masses and orbital periods to evolve in the future. The masses and
orbital periods of these systems were taken from the catalogue of
\cite{2001NewAR..45..135V}, to which we refer for the original references 
for these systems. We have calculated the future evolution of
these systems semi-empirically, as follows. We adopted the Conti-scenario \citep{Conti1976}
for the evolution of WR stars. According to this model, WR stars begin as 
WN stars and then evolve with strong wind mass loss into WC/WO stars, 
that finally undergo core collapse \citep{2007ARA&A..45..177C}. 
(For an alternative point of view, see \cite{2012A&A....540..144V,2016MNRAS.459..1505}).

To calculate the evolution of the observed WR+O binaries in Table 1,
we made the following assumptions:

\begin{itemize}
\item On the basis of evolution calculations, such as by the Geneva group, 
  we assumed, adopting the Conti scenario, the WR stars
  to spend 70\% of their helium-burning lifetime as a WN star and 30\%
  as a WC star.

\item We assumed an observed WN star to be half-way its WN lifetime,
  that is: at 35\% of its helium star lifetime. Similarly, we assumed
  observed WC stars to be at 85\% of their helium star lifetime, so
  still to have 15\% of this lifetime to go.

\item We assumed their wind mass loss rates over their entire lifetime
  to correspond to the ones stars of their presently observed WR
  type. For these rates we used wind mass loss rates 1.4 times lower
  than assumed by \cite{1992A&AS...96..269S}, for the following
  reasons: Schaller et al. found for their adopted wind mass loss
  rates for solar metallicity that even stars up to 120\MSun
  finished with a mass of only 8\MSun. It is evident that this
  cannot be correct, since the Population I X-ray binary Cygnus X-1 is
  a 14.8\MSun black hole(\cite{2011ApJ...742...840},\cite{2012PoS..054....V}). 
  Assuming some 90 per cent of the final
  WC star (basically a CO core) to become the black hole (\cite[e.g.][]{2012ASSL..384..299H}), the final mass of the WR progenitor of Cyg X-1 must have been
  16.4\MSun. The formation of a black hole with the mass of Cyg X-1
  is possible only if the real wind mass loss rates at solar
  metallicity are about 1.4 times lower than the ones used by Schaller
  et al. In that case the progenitor of Cyg X-1 was a star of about
  80\MSun with solar metallicity.

\item Also for the O-stars we used wind loss rates of 1.4 times lower
  than the ones used by \cite{1992A&AS...96..269S}.

\item As total lifetime of the WR stars (massive helium stars) we used
  400 000 yrs \citep{2015MNRAS..449..2322P}.
  \end{itemize}

The results of these calculations are listed in Table 1.
Columns 2, 3 and 4 list the observed parameters of the
systems. Column 5 lists the masses of the components after the WR star
has terminated its evolution and has become a black hole, and
subsequently the companion has evolved for 3 million years to leave
the main sequence and start Roche-lobe overflow. We calculated the
masses in column 5 by taking into account the wind mass-loss of both
stars, and by assuming that at the end of the life of the Wolf-Rayet
star as a WC-type star, 90\% of the mass of the WC star disappeared
into the black hole \citep[cf.][]{2012ASSL..384..299H,2016ApJ...821...38S}, meaning that 
just only the gravitational binding energy of the helium star is lost. As there is no mass lost, we have assumed that the black holes did not receive a natal kick.
(We realize that our implicit assumption in these calculations that stars with initial masses larger than about 25\MSun always leave black holes, may not be fully realistic, since \citet{2016ApJ...821...38S} (and also Ertl et al. and Muller in 2015, in papers referred to in this paper) find from their calculations that, quite arbitrarily, sometimes even very massive stars may still leave a neutron star as a remnant. Since the precise reasons why this happens is not fully understood, we have chosen not to take this into account here). 

A possible indication that WR stars tend to leave black holes is that, so far, no WR stars have been detected as progenitors of Type Ib,c supernovae \citep{2015PASA...32...16S}. 

We calculated how the orbits
changed due to the wind mass loss and the core-collapse gravitational-mass loss, and
assumed that after this the orbits tidally synchronized again. We then
assumed the O-type companions of the BHs still to live another 3
million years as core-hydrogen burning stars, losing mass by stellar
wind in this period. For the top six systems in the table, we subsequently calculated the spiral in of the
resulting HMXB consisting of the O-star and the BH, using the
SS433-type of spiral in. This resulted into the
WR+BH systems with masses and orbital periods listed in columns (6) and
(7), respectively. For the 7th system, the mass ratio of O-star and black hole is larger than 3.5 and we assumed this system to pass through CE-evolution. 

One notices that several of the top-six WR X-ray binaries
have orbital periods of order 1 to 2 days, and that their compact
stars are black holes. After taking account of the wind mass loss of
the WR stars in the WR X-ray systems during their further evolution and,
again assuming that 90 per cent of the mass of the final WC star
disappears into the black hole, one obtains masses and orbital periods
of the double black hole binaries resulting from these systems listed
in columns 8 and 9, respectively. (We assumed that the wind mass loss did not increase the orbital period of the WR X-ray binary, as the gravitational torque excerted by the black hole on the outflowing thick wind will cause loss of orbital angular momentum, which is expected to compensate the increase of orbital period due to the mass loss; if the latter torques would not occur, the final orbital periods will be about 1.8 times larger). 

One observes from this table that
some well-known WR+O spectroscopic binaries with orbital periods of
order one week can produce close WR+BH X-ray binaries, and
subsequently produce close double black hole binaries. 

Since the first 
three systems in Table~\ref{Table1} have a mass ratio of donor and black hole
before spiral-in mass transfer very close to the 3.5 limit, it is not
completely sure whether the mass transfer from the radiative envelope
will not become unstable.  To mark this uncertainty, we have a put a
colon after the post-spiral-in orbital periods in Table~\ref{Table1}.
For the 4th to 6th systems in Table~\ref{Table1} the mass ratio before
spiral-in is below 3.5 and the transfer is expected to certainly be
stable. The 4th and 5th systems, however, do not leave close WR X-ray binaries,
and also not close double black holes. They leave WR X-ray binaries resembling 
the one observed system with an orbital period of about 8 days, mentioned in the Introduction. 
(In the VIIth catalogue of WR stars \citep{2001NewAR..45..135V} there are 3 more systems (WR30, 113 and 141) that will evolve similarly and leave wide WR X-ray binaries and wide double BHs). 

In the case of the first three
systems in Table~\ref{Table1}: if their mass transfer becomes unstable, a Common
Envelope will form and the outcome of the evolution must be calculated
using equation~\ref{Eq:af_ai}. It turns out that with the above
assumed value of $\alpha_{\rm CE}\,\lambda = 0.25 - 0.45$, none of these three
systems survives the CE-evolution, as in each case the Roche-lobe
radius of the post-CE helium star is smaller than the radius of this
star. 
\begin{table*}
  \caption{Anticipated future evolution of seven well-observed WR+O
    binaries. The orbital periods, spectral types and 
    masses of the components were taken from the catalogue by \citet{2001NewAR..45..135V}. 
    The WR stars and O-stars in
    these systems are sufficiently massive to leave black hole
    remnants. The ways how the masses of these black holes (bold-print
    numbers) were calculated are explained in the text. It is assumed for the top six systems
    that the resulting O+BH systems will spiral-in following the
    SS433-recipe. In the three first systems, the mass ratio of O-donor
    and BH is very close to 3.5, such that it is not fully certain that
    the mass transfer will be stable (see text). To indicate this
    uncertainty, a colon was placed after the short orbital periods of
    the resulting WR-XRBs. Assuming the SS433-type mass transfer indeed to be
    stable, the masses of the double black holes, with their orbital periods,
    indicated in the last columns, will result. The very last column lists the GW-merger times of these systems (no times are given if the merger time is longer than the age of the Universe). In case the SS433-like mass transfer would not be stable, the WR/X-ray systems indicated with colons will go into CE evolution; in that case none of these systems will survive. The bottom line gives the anticipated evolution WR 11(Gamma-2 Velorum), which will in the second phase of mass transfer go into Common Envelope evolution and produce a very short-period WR/X binary and double black hole}
  \label{Table1}
  \begin{center}
  \begin{tabular}{l|ccccccccc}
  \hline
       &          & \multicolumn{2}{c}{Observed} & HMXB at RLOF & \multicolumn{2}{c}{WR X-ray binary}& \multicolumn{3}{c}{Double black hole} \\
  Name & Spectrum & $P$     & Masses   & Masses & Masses     & $P$     & Masses & $P$ & $t(merge)$ \\
       &          & [day]   & [\MSun]  & [\MSun]& [\MSun]    & [day]   & [\MSun]& [day]  & [Gyr]\\
  \hline
  WR 127&WN3 + O9.5V&   9.555 & 17 + 36   &{\bf 9.6} + 33& {\bf 9.6} + 13.6 &  1.54:   & {\bf 9.6} + {\bf 7.0}& 1.71 & 6.91\\
  WR  21&WN5 + O4-6&   8.255 & 19 + 37   &{\bf 10.2} + 34& {\bf 10.2} + 14.1&  1.64:   &{\bf 10.2} + {\bf 7.9}& 1.77 & 6.50\\
  WR 62a&WN5 + O5.5&    9.145& 22 + 40.5 &{\bf 10.8} + 37& {\bf 10.8} + 16.1 &  1.45:    &{\bf 10.8} + {\bf 8.8}& 1.61 & 4.40\\
  WR  42&WC7 + O7V &    7.886& 14 + 23   &{\bf 10.4} + 22& {\bf 10.4} + 8.0 & 13.75    &{\bf 10.4} + {\bf 4.5}& 14.71& --\\
  WR  47&WN6 + O5V &   6.2393& 51 + 60  &{\bf 18.1} + 46 & {\bf 18.1} + 25.8 &  7.96    &{\bf 18.1} + {\bf 10.4}& 8.59 & --\\ 
  WR 79&WC7 + O5-8 &   8.89&   11 + 29  &{\bf  9.0}  + 27.4& {\bf 9.0}  + 10.1 &  2.44   &{\bf 9.0} + {\bf 5.4}&  2.64 & -- \\
  WR 11(CE)&WC8 + O7.5III& 78.53& 9.0 + 30&{\bf 7.8} + 28.5& {\bf 7.8} + 10.5 & 0.90 &{\bf 7.8} +{\bf 5.6}& 0.98 & 2.24\\
  \hline
  \end{tabular}
  \end{center}
\end{table*}

One sees that the orbital periods of WR XRBs produced by stable Roche
lobe overflow are considerably larger than the one produced by CE evolution in the system  of WR11.

\section{The galactic formation rate of WR X-ray binaries}

\subsection{Predicted galactic numbers of WR X-ray binaries, compared to observations: still too few observed systems}

\cite{2015MNRAS..449..2322P} estimate the number of WR stars in the Galaxy to be $1200(\pm 200)$. WR X-ray binaries have in our model come from WR plus O-type binaries similar to the systems in Table 1. In these systems the WR star and the O-star have comparable luminosities, meaning that they are double-lined spectroscopic binaries (abbreviated as SB2), which started out as O-type binaries with components with roughly similar masses. \cite{2001NewAR..45..135V} found 13 of the 61 WR stars within 3 kpc from the sun to  be SB2s, so about 20 per cent, which would imply 240 SB2 WR-systems in the Galaxy. We particularly selected the 7 systems in Table 1, plus the 3 other systems mentioned in the foregoing section, for their masses and orbital periods such that they could potentially produce WR X-ray binaries. They therefore are not a representative sample of the SB2  WR+O binaries in van der Hucht's VIIth catalogue of WR stars, in which there are in total 31 SB2 systems. Assuming these ten systems to be representative for one third of all WR SB2 systems in the Galaxy, then 80 such systems may be expected to evolve similarly to the ten systems mentioned in the last section, and produce similar WR/X-ray binaries. As in WR+O binaries and in WR X-ray binaries the WR stars are expected to live equally long, one would then expect, in a steady state of star formation, that there are also 80 WR X-ray binaries in the Galaxy. However, we know only one such system in the Galaxy: Cygnus X-3, which has an orbital period of 0.2 days. This system, at some 8-10 kpc distance, is in absolute terms one of the brightest X-ray sources in the Galaxy; although only our side of the Galaxy   has been well-surveyed for X-ray sources, it is unlikely that there are more than 3 similar systems in the entire Galaxy. The large absolute X-ray luminosity of Cyg X-3 may be due to its short orbital period, which implies that the compact star is moving in the low-velocity part of its stellar wind. As the accretion rate scales with the minus 4th power of the wind velocity times the minus 2nd power of the orbital radius, in systems with orbital periods of one day or more, like in Table 1, the compact stars are in the high-velocity part of the wind, and are at some 3 times larger orbital radius, such that their X-ray luminosities will be at least some three orders of magnitude lower than that of Cyg X-3, i.e. below $\sim 10^{35}$\ ergs/s. Since, however, we know none of such systems among the well-surveyed part of our Galaxy, up to some 6 kpc distance, which is one-fifth of the area of the galactic disk, the number in the entire galaxy is unlikely to be larger than about 5. So, the total number of observable WR X-ray binaries in the Galaxy is unlikely to be larger than 8, which is at most only 10 per cent of the above expected number. In the following section we discuss possible ways out of this conundrum.
\subsection{Conceivable explanations for the large difference between expected and observed numbers of galactic black-hole WR X-ray binaries}
To calculate the parameters of the WR X-ray binaries in Table 1 a number of assumptions about the evolution of the WR+O binary predecessors were made, as described in section 3.4. We now critically examine the effects of changing a number of the underlying assumptions.
\begin{itemize}
\item (i) The assumption that for $q < 3.5$ the systems evolve with stable Roche-lobe overflow. If the real q-limit would be lower, e.g. $q < 3$, the six uppermost O + BH systems in Table 1 will all go into CE-evolution and will not survive. Also the three systems WR30, 113 and 141 mentioned in section 3.4 will not survive. Only the system WR 11 with its large orbital period will, going through CE-evolution, survive. This could reduce the expected number of WR X-ray binaries by a factor of 10. It would then imply that there are some 8 WR X systems in the Galaxy with orbital periods like that of the descendent of WR 11: about one day.  

\item (ii) The assumption that the BH formed from a WC star has 90 per cent of the final mass of this star. \citet{1999A&A...352L..87N} argue that only 0.65 of the final mass of a helium star ends up in the BH. We repeated the calculations of the evolution of the systems in Table 1 with this assumption. The first three systems in Table 1, as well as WR 79 will now go into CE evolution and coalesce, while WR 42 and 47 go through "isotropic re-emission" and terminate with periods of about 4 days; WR 30, 113 and 141 evolve in a similar way and produce systems with still longer periods. The only system that in this case still survives as a short-period WR X-ray binary is WR11, which terminates with an orbital period of 2.13 hours if $\alpha_{\rm CE} = 0.50$ and 4.98 hours if $\alpha_{\rm CE} = 0.90$.

\item (iii) Our assumption that the core collapses of the massive stars always leave black holes. As mentioned in section 3.4, the results of the core-collapse calculations of massive stars by \citet{2016ApJ...821...38S} show that, while sometimes even stars in the mass range 15 to 20\MSun may leave a black hole, also in very massive stars the core collapse may in a number of cases still lead to the formation of a neutron star. If this is indeed the case, and would, for example, occur in half of the cases of the core collapses involved in the systems in Table 1, it would reduce the resulting number of WR X-ray binaries produced by these systems by a factor 2: if half of the systems produced a neutron star in the first core collapse in the system, these systems would merge on Roche-lobe overflow. And the produced number of double black holes would be reduced by a factor 4, as also in the second supernova half of the collapses would produce a neutron star.    
\end{itemize}

\section{Discussion - possible galactic formation rate of double black holes originating from WR X-ray binaries}

From the discussion in the last section, it appears that lowering the fraction of the final WR-star mass that goes to form the black hole does not solve the problem of the lack of Galactic WR X-ray binaries. On the other hand, we found that a more promising way to explain the low observed incidence of WR X-ray binaries is: a lower q-limit than $ q=3.5 $ for stable mass transfer by Roche-lobe overflow. The latter limiting value was derived for binary systems with components less massive than 12 solar masses. It is therefore crucial to make more detailed
binary evolution calculations for more massive black-hole HMXBs, in order to more precisely estimate this q-limit for  massive systems (see also \citet{2017MNRAS.465.2092P}). 

Also, it is important to calculate this limit for different values of the metallicity, as this type of evolution is expected to have been important also in the early universe. Further, we saw that if a sizeable fraction of the core-collapses of massive stars would still leave a neutron star instead of a black hole, this could further reduce the produced number of WR X-ray binaries by a factor two. This would then result in only 4 such system sin the Galaxy at any time. 

Finally, assuming that the solution of the low incidence of WR X-ray binaries in the galaxy is indeed a lower q-limit than 3.5, and combining this with half of the massive star core collapses producing neutron stars, one can make an estimate of the galactic formation and merger rate of close double black holes resulting from WR X-ray binaries. As mentioned above, the produced double black hole systems then resemble the systems resulting from WR11, which have orbital periods of around one day. Such systems will merge on a timescale of order $\sim10^{9}$\ yr by the loss of gravitational radiation. As we expect half of the estimated 4 close WR X-ray binaries to form a close double black hole in 400 000 yr (the WR lifetime), the galactic merger rate of close double black holes produced by this process is  $\simeq 0.5\cdot 10^{-5}$\ per yr. The rate based on the system of Cygnus X-3 alone - assuming its WR star to be massive enough to leave a black hole- is of similar order, as its WR lifetime is again of order 400 000 years (see also \citet{2015MNRAS..482..1112D} and references therein). Although the uncertainties on these estimated rates are expected to be at least an order of magnitude (each way), they nevertheless are interesting.  The rates are an order of magnitude lower than the estimated galactic double-neutron-star merger rate \citep{2004ApJ...614L.137K}.  

\section*{Acknowledgments}

We thank Thomas Tauris for very useful comments.  This work was
supported in part by the National Science Foundation under Grant No. NSF PHY11-25915. The CHARM is project P7/08 in the phase VII IA. It further was supported by the Netherlands Research Council NWO (grants
\#643.200.503, \#639.073.803 and \#614.061.608) by the Netherlands
Research School for Astronomy (NOVA). 
SdM acknowledges support by a Marie Sklodowska-Curie Action (H2020 MSCA-IF-2014, project id 661502). Part of the numerical
computations were carried out on the Little Green Machine at Leiden
University (612.071.305 and 621.016.701).

%\input /home/spz/Latex/lib/bib/references
%\input journals.tex
%\input spz.bbl

%\bibliographystyle{mn2e}
%\bibliography{SS433}

\begin{thebibliography}{}
\bibitem[\protect\astroncite{{Begelman} et~al.}{2006}]{2006MNRAS.370..399B}
{Begelman}, M.~C., {King}, A.~R., {Pringle}, J.~E. 2006, \mnras, 370, 399

\bibitem[\protect\astroncite{{Belczynski} et~al.}{2016}]{2016Natur.534..512B}
{Belczynski}, K., {Holz}, D.~E., {Bulik}, T., {O'Shaughnessy}, R. 2016, \nat,
  534, 512

\bibitem[\protect\astroncite{{Bhattacharya} \& {van den
  Heuvel}}{1991}]{1991PhR...203....1B}
{Bhattacharya}, D., {van den Heuvel}, E.~P.~J. 1991, \physrep, 203, 1

\bibitem[\protect\astroncite{{Casares} et~al.}{2014}]{2014Natur.505..378C}
{Casares}, J., {Negueruela}, I., {Rib{\'o}}, M., {Ribas}, I., {Paredes}, J.~M.,
  {Herrero}, A., {Sim{\'o}n-D{\'{\i}}az}, S. 2014, \nat, 505, 378

\bibitem[\protect\astroncite{{Clayton}}{1968}]{1968psen.book.....C}
{Clayton}, D.~D. 1968,
\newblock {Principles of stellar evolution and nucleosynthesis}

\bibitem[\protect\astroncite{{Conti}}{1976}]{Conti1976}
{Conti}, P. 1976, Mem Soc Roy Sciences Li\`ege 6e S\'er., Tome IX, p.\,193

\bibitem[\protect\astroncite{{Crowther}}{2007}]{2007ARA&A..45..177C}
{Crowther}, P.~A. 2007, \araa, 45, 177

\bibitem[\protect\astroncite{{de Kool}}{1990}]{1990ApJ...358..189D}
{de Kool}, M. 1990, \apj, 358, 189

\bibitem[\protect\astroncite{{de Mink} \& {Mandel}}{2016}]{2016MNRAS.460.3545D}
{de Mink}, S.~E., {Mandel}, I. 2016, \mnras, 460, 3545

\bibitem[\protect\astroncite{{Dewi} et~al.}{2005}]{2005MNRAS.363L..71D}
{Dewi}, J.~D.~M., {Podsiadlowski}, P., {Pols}, O.~R. 2005, \mnras, 363, L71

\bibitem[\protect\astroncite{{Eggleton}}{1983}]{1983ApJ...268..368E}
{Eggleton}, P.~P. 1983, \apj, 268, 368

\bibitem[\protect\astroncite{{Ekstr{\"o}m} et~al.}{2012}]{2012A&A...537A.146E}
{Ekstr{\"o}m}, S., {Georgy}, C., {Eggenberger}, P., {Meynet}, G., {Mowlavi},
  N., {Wyttenbach}, A., {Granada}, A., {Decressin}, T., {Hirschi}, R.,
  {Frischknecht}, U., {Charbonnel}, C., {Maeder}, A. 2012, \aap, 537, A146
  
\bibitem[\protect\astroncite{{Eldridge} \& {Stanway}}{2016}]{2016MNRAS..462..3302E}
{Eldridge}, J.J., {Stanway}, E.R., \mnras, 462, 3302
  
\bibitem[\protect\astroncite{{Esposito} et~al.}{2015}]{2015MNRAS..482..1112D}
{Esposito}, P., {Israel}, G.~L., {Milisavljevic}, D., {Mapelli}, M., {Zampieri}, L., {Sidoli}, L., {Rodriguez Castillo}, G.~A. 2015, \mnras, 482, 1112

\bibitem[\protect\astroncite{{Fabrika}}{2004}]{2004ASPRv..12....1F}
{Fabrika}, S. 2004, Astrophysics and Space Physics Reviews, 12, 1

\bibitem[\protect\astroncite{{Grudzinska} et~al.}{2015}]{2015MNRAS.452.2773G}
{Grudzinska}, M., {Belczynski}, K., {Casares}, J., {de Mink}, S.~E.,
  {Ziolkowski}, J., {Negueruela}, I., {Rib{\'o}}, M., {Ribas}, I., {Paredes},
  J.~M., {Herrero}, A., {Benacquista}, M. 2015, \mnras, 452, 2773

\bibitem[\protect\astroncite{{Heger}}{2012}]{2012ASSL..384..299H}
{Heger}, A. 2012,
\newblock in K. {Davidson}, R.~M. {Humphreys} (eds.), Eta Carinae and the
  Supernova Impostors, Vol. 384 of {\em Astrophysics and Space Science
  Library\/},  299

\bibitem[\protect\astroncite{{Hillwig} \& {Gies}}{2008}]{2008ApJ...676..L37H}
{Hillwig}, T.~C., {Gies}, D.~R. 2008, \apjl, 676, L37

\bibitem[\protect\astroncite{{Humphreys} \&
  {Davidson}}{1994}]{1994PASP..106.1025H}
{Humphreys}, R.~M., {Davidson}, K. 1994, \pasp, 106, 1025

\bibitem[\protect\astroncite{{Hurley} et~al.}{2000}]{2000MNRAS.315..543H}
{Hurley}, J.~R., {Pols}, O.~R., {Tout}, C.~A. 2000, \mnras, 315, 543

\bibitem[\protect\astroncite{{Ivanova} et~al.}{2013}]{2013A&ARv..21...59I}
{Ivanova}, N., {Justham}, S., {Chen}, X., {De Marco}, O., {Fryer}, C.~L.,
  {Gaburov}, E., {Ge}, H., {Glebbeek}, E., {Han}, Z., {Li}, X.-D., {Lu}, G.,
  {Marsh}, T., {Podsiadlowski}, P., {Potter}, A., {Soker}, N., {Taam}, R.,
  {Tauris}, T.~M., {van den Heuvel}, E.~P.~J., {Webbink}, R.~F. 2013, \aapr,
  21, 59

\bibitem[\protect\astroncite{{Kalogera} \& {Baym}}{1996}]{1996ApJ...470L..61K}
{Kalogera}, V., {Baym}, G. 1996, \apjl, 470, L61

\bibitem[\protect\astroncite{{Kalogera} et~al.}{2004}]{2004ApJ...614L.137K}
{Kalogera}, V., {Kim}, C., {Lorimer}, D.R., {Burgay}, M., {D'Amico}, N., {Possenti}, A., {Manchester}, R.N., {Lyne}, A.G., {Joshi}, B.C., {McLaughlin}, M.A., {Kramer}, M., {Sarkissian}, J.M., {Camilo}, F. 2004, \apjl, 614, L137

\bibitem[\protect\astroncite{{King} \& {Begelman}}{1999}]{1999ApJ...519L.169K}
{King}, A.~R., {Begelman}, M.~C. 1999, \apjl, 519, L169

\bibitem[\protect\astroncite{{King} et~al.}{2000}]{2000ApJ...530L..25K}
{King}, A.~R., {Taam}, R.~E., {Begelman}, M.~C. 2000, \apjl, 530, L25

\bibitem[\protect\citeauthoryear{Kippenhahn \& Weigert}{1990}]{1990sse..book.....K} Kippenhahn R., Weigert A., 1990, sse..book, 192 

\bibitem[\protect\astroncite{{Kruckow} et~al.}{2016}]{2016A&A..596...A58K}
{Kruckow}, M.~U., {Tauris}, T.~M., {Langer}, N., {Szecsi}, D., {Marchant}, P., {Podsiadlowski}, Ph. 2016,\aap, 596, A58

\bibitem[\protect\astroncite{{Lipunov} et~al.}{1997}]{1997AstL...23..492L}
{Lipunov}, V.~M., {Postnov}, K.~A., {Prokhorov}, M.~E. 1997, Astronomy Letters,
  23, 492
  
\bibitem[\protect\astroncite{{Lommen} et~al.}{2005}]{2005A&A...443..231L}
{Lommen}, D., {Yungelson}, L., {van den Heuvel}, E. P. J.,{Nelemans}, G., Portegies Zwart, S. 2005, \aap, 443, 231

\bibitem[\protect\astroncite{{Marchant} et~al.}{2016}]{2016A&A...588A..50M}
{Marchant}, P., {Langer}, N., {Podsiadlowski}, P., {Tauris}, T.~M., {Moriya},
  T.~J. 2016, \aap, 588, A50
  
\bibitem[\protect\astroncite{{Massevitch} \& {Yungelson}}{1975}]{1975MmSAI..46..217M}
{Massevitch}, A., {Yungelson}, L. 1975, \memsai, 46, 217
  
\bibitem[\protect\astroncite{{McClelland} \& {Eldridge}}{2016}]{2016MNRAS.459..1505}
{McClelland}, L.A.S., {Eldridge}, J.J. 2016, \mnras, 459, 1505

\bibitem[\protect\astroncite{{Nauenberg} \&
  {Chapline}}{1973}]{1973ApJ...179..277N}
{Nauenberg}, M., {Chapline}, Jr., G. 1973, \apj, 179, 277

\bibitem[\protect\astroncite{{Nelemans} et~al.}{1999}]{1999A&A...352L..87N}
{Nelemans}, G.,{Tauris}, T.M., {van den Heuvel}, E.P.J. 1999, \aap, 352, L87 

\bibitem[\protect\astroncite{{Orosz} et~al.}{2011}]{2011ApJ...742...840}
{Orosz}, J.~A., {McClintock}, J.~E., {Aufdenberg}, J.~P., {Remillard}, R.~A.,
  {Reid}, M.~J., {Narayan}, R., {Gou}, L. 2011, \apj, 742, 84

\bibitem[\protect\astroncite{{Owocki}}{2015}]{2015ASSL..412..113O}
{Owocki}, S.~P. 2015,
\newblock in J.~S. {Vink} (ed.), Very Massive Stars in the Local Universe, Vol.
  412 of {\em Astrophysics and Space Science Library\/},  113

\bibitem[\protect\astroncite{{Paczy{\'n}ski}}{1967}]{1967AcA....17..355P}
{Paczy{\'n}ski}, B. 1967, \actaa, 17, 355

\bibitem[\protect\astroncite{{Pavlovskii} et~al.}{2017}]{2017MNRAS.465.2092P}{Pavlovskii}, K.,{Ivanova}, N., {Belczynski}, K.,{Van}, K.X. 2017, \mnras, 465, 2092

\bibitem[\protect\astroncite{{Portegies Zwart}}{2013}]{2013MNRAS.429L..45P}
{Portegies Zwart}, S. 2013, \mnras, 429, L45

\bibitem[\protect\astroncite{{Portegies Zwart} \&
  {Verbunt}}{1996}]{1996A&A...309..179P}
{Portegies Zwart}, S.~F., {Verbunt}, F. 1996, \aap, 309, 179

\bibitem[\protect\citeauthoryear{Portegies Zwart \& Spreeuw}{1996}]{1996A&A...312..670P} Portegies Zwart S.~F., Spreeuw H.~N., 1996, A\&A, 312, 670 

\bibitem[\protect\astroncite{{Raguzova} \& {Popov}}{2005}]{2005A&AT...24..151R}
{Raguzova}, N.~V., {Popov}, S.~B. 2005, Astronomical and Astrophysical
  Transactions, 24, 151

\bibitem[\protect\astroncite{{Reed}}{2000}]{2000AJ....120..314R}
{Reed}, B.~C. 2000, \aj, 120, 314

\bibitem[\protect\astroncite{{Reig}}{2011}]{2011Ap&SS.332....1R}
{Reig}, P. 2011, \apss, 332, 1

\bibitem[\protect\astroncite{{Roberts}}{1957}]{1957PASP...69..59}
{Roberts}, W.~S. 1957, \pasp, 69, 59

\bibitem[\protect\astroncite{{Rosslowe} \& {Crowther}}{2015}]{2015MNRAS..449..2322P}
{Rosslowe}, C.~K., {Crowther}, P.~A. 2015, \mnras, 449, 2322

\bibitem[\protect\astroncite{{Sana} et~al.}{2012}]{2012Sci...337..444S}
{Sana}, H., {de Mink}, S.~E., {de Koter}, A., {Langer}, N., {Evans}, C.~J.,
  {Gieles}, M., {Gosset}, E., {Izzard}, R.~G., {Le Bouquin}, J.-B.,
  {Schneider}, F.~R.~N. 2012, Science, 337, 444
  
\bibitem[\protect\astroncite{{Sander} et~al.}{2012}]{2012A&A....540..144V}
{Sander}, A., {Hamann}, W.~-R., {Todt}, H. 2012, \aap, 540, 144  

\bibitem[\protect\astroncite{{Schaller} et~al.}{1992}]{1992A&AS...96..269S}
{Schaller}, G., {Schaerer}, D., {Meynet}, G., {Maeder}, A. 1992, \aaps, 96, 269

\bibitem[\protect\astroncite{{Shakura} \&
  {Sunyaev}}{1973}]{1973A&A....24..337S}
{Shakura}, N.~I., {Sunyaev}, R.~A. 1973, \aap, 24, 337

\bibitem[\protect\astroncite{{Soberman} et~al.}{1997}]{1997A&A...327..620S}
{Soberman}, G.~E., {Phinney}, E.~S., {van den Heuvel}, E.~P.~J. 1997, \aap,
  327, 620
  
\bibitem[\protect\astroncite{{Smartt}}{2015}]{2015PASA...32...16S} {Smartt}, S.J. 2015, \pasa, 32, 16 
  
\bibitem[\protect\astroncite{{Sukhbold} et~al.}{2016}]{2016ApJ...821...38S}
{Sukhbold}, Tuguldur, {Ertl}, T., {Woosley}, S.~E., {Brown}, J.~M., Janka, H.~-T. 2016, \apj, 821, 38

\bibitem[\protect\astroncite{{Taam}}{1996}]{1996IAUS..165....3T}
{Taam}, R.~E. 1996,
\newblock in J. {van Paradijs}, E.~P.~J. {van den Heuvel}, E. {Kuulkers}
  (eds.), Compact Stars in Binaries, Vol. 165 of {\em IAU Symposium\/}, ~3

\bibitem[\protect\astroncite{{Taam} \& {Sandquist}}{2000}]{2000ARA&A..38..113T}
{Taam}, R.~E., {Sandquist}, E.~L. 2000, \araa, 38, 113

\bibitem[\protect\astroncite{{Tauris}}{1996}]{1996A&A...315..453T}
{Tauris}, T.~M.1996, \aap, 315, 453

\bibitem[\protect\astroncite{{Tauris} et~al.}{2000}]{2000ApJ...530L..93K}
{Tauris}, T.~M.,{van den Heuvel}, E.~P.~J.,{Savonije}, G.~J.2000\apj, 530, L93

\bibitem[\protect\astroncite{{Tauris} \& {van den
  Heuvel}}{2006}]{2006csxs.book..623T}
{Tauris}, T.~M., {van den Heuvel}, E.~P.~J. 2006,
\newblock {Formation and evolution of compact stellar X-ray sources},  623--665

\bibitem[\protect\astroncite{{Tout} et~al.}{1997}]{1997MNRAS.291..732T}
{Tout}, C.~A., {Aarseth}, S.~J., {Pols}, O.~R., {Eggleton}, P.~P.1997\mnras, 291, 732

\bibitem[\protect\astroncite{{Tutukov} \&
  {Yungelson}}{1973}]{1973NInfo..27...70T}
{Tutukov}, A., {Yungelson}, L. 1973, Nauchnye Informatsii, 27, 70

\bibitem[\protect\astroncite{{Tutukov} \&
  {Yungelson}}{1993}]{1993MNRAS.260..675T}
{Tutukov}, A.~V., {Yungelson}, L.~R. 1993, \mnras, 260, 675

\bibitem[\protect\citeauthoryear{Vanbeveren, van Rensbergen, \& De Loore}{1982}]{1982A&A...115...69V} Vanbeveren D., van Rensbergen W., De Loore C., 1982, A\&A, 115, 69 

\bibitem[\protect\astroncite{{van den Heuvel}}{1994}]{1994inbi.conf..263V}
{van den Heuvel}, E.~P.~J. 1994,
\newblock in S.~N. {Shore}, M. {Livio}, E.~P.~J. {van den Heuvel}, H.
  {Nussbaumer}, A. {Orr} (eds.), Saas-Fee Advanced Course 22: Interacting
  Binaries, p.~263

\bibitem[\protect\astroncite{{van den Heuvel} \& {De
  Loore}}{1973}]{1973A&A....25..387V}
{van den Heuvel}, E.~P.~J., {De Loore}, C. 1973, \aap, 25, 387

\bibitem[\protect\astroncite{{van der Hucht}
  et~al.}{2003}]{2003IAUS..212.....V}
{van der Hucht}, K., {Herrero}, A., {Esteban}, C. (eds.) 2003,
\newblock {A massive star odyssey : from main sequence to supernova :
  proceedings of the 212th Symposium of the International Astronomical Union
  held in Costa Teguise, Lanzarote, Canary Islands, 24-28 June 2002}, Vol. 212
  of {\em IAU Symposium\/}

\bibitem[\protect\astroncite{{van der Hucht}}{2001}]{2001NewAR..45..135V}
{van der Hucht}, K.~A. 2001, \nar, 45, 135

\bibitem[\protect\astroncite{{van Kerkwijk} et~al.}{1992}]{1992Natur.355..703V}
{van Kerkwijk}, M.~H., {Charles}, P.~A., {Geballe}, T.~R., {King}, D.~L.,
  {Miley}, G.~K., {Molnar}, L.~A., {van den Heuvel}, E.~P.~J., {van der Klis},
  M., {van Paradijs}, J. 1992, \nat, 355, 703

\bibitem[\protect\astroncite{{Webbink}}{1984}]{1984ApJ...277..355W}
{Webbink}, R.~F. 1984, \apj, 277, 355

\bibitem[\protect\astroncite{{Ziolkowski}}{2012}]{2012PoS..054....V}
{Ziolkowski}, J.2012, \pos, 054

\end{thebibliography}
%\input SS433.bbl

\end{document}